# Data-Driven Machine Learning to Predict Mechanical Properties of Monolayer TMDs

Prottay Malakar[1], Md Shajedul Hoque Thakur[2], Shahriar Muhammad Nahid[3] and Md Mahbubul Islam[4*]

[1]*Department of Mechanical Engineering, Bangladesh University of Engineering & Technology (BUET), Dhaka, 1000, Bangladesh*

[2]*Department of Materials Science and NanoEngineering, Rice University, Houston, TX 77005, United States*

[3]*Department of Mechanical Science and Engineering, University of Illinois Urbana Champaign, Urbana, IL 61801, United States*

[4]*Department of Mechanical Engineering, Wayne State University, 5050 Anthony Wayne Drive, Detroit, MI 48202, United States*

[*]Corresponding author, *Email address:* gy5553@wayne.edu

# Abstract

The understanding of the material properties of the layered transition metal dichalcogenides (TMDs) is critical for their applications in structural composites. The data-driven machine learning (ML) based approaches are being developed in contrast to traditional experimental or computational approach to predict and understand materials properties under varied operating conditions. In this study, we used two ML algorithms such as Long Short-Term Memory (LSTM) and Feed Forward Neural Network (FFNN) combined with molecular dynamics (MD) simulations to predict the mechanical properties of $MX_2$ (M=Mo, W, and X=S, Se) TMDs. The LSTM model is found to be capable of predicting the entire stress-strain response whereas the FFNN is used to predict the material properties such as fracture stress, fracture strain, and Young's modulus. The effects of operating temperature, chiral orientation, and pre-existing crack size on the mechanical properties are thoroughly investigated. We carried out 1440 MD simulations to produce the input dataset for the neural network models. Our results indicate that both LSTM and FFNN are capable of predicting the mechanical response of monolayer TMDs under different conditions with more than 95% accuracy. The FFNN model exhibits lower computational cost than LSTM; however, the capability of LSTM model to predict the entire stress-strain curve is advantageous to assess material properties. The study paves the pathway toward extending this approach to predict other important properties, such as optical, electrical, and magnetic properties of TMDs.

# Introduction

Two-dimensional (2D) transition-metal dichalcogenides (TMDs) have opened new possibilities to realize extraordinary functionalities for next-generation electronic, optoelectronic, and quantum devices[1], [2]. The new and superior functions of these materials are derived from the quantum confinement effect of free charges in the out-of-the-plane direction, the existence of a direct bandgap in their monolayer regime, their transparent and flexible nature, and their extraordinary and unusual mechanical properties[1]–[7]. Monolayer TMDs such as $MoS_2$, $MoSe_2$, $WS_2$ and $WSe_2$ have emerged as potential candidates for flexible electronic/optoelectronic applications and nanoelectromechanical systems (NEMS) devices[8], [9]. TMDs have potential use as nano resonators[10], [11], piezoelectric devices[12]–[15], vibrational energy harvesters[16], [17], optical modulators[18], and in a variety of electrical, optoelectronic, and energy devices[19]–[23]. For many of these applications, the materials are exposed to a variety of operating conditions, thus understanding of their mechanical behaviour in various conditions is crucial to assuring their continued use in these devices.

Both computational and experimental methods have been widely used to investigate the mechanical characteristics of TMDs[24]–[28]. Density functional theory (DFT) and molecular dynamics (MD) are the mostly used computational techniques to investigate the mechanical properties and fracture mechanisms of TMDs[24], [28]–[30]. For instance, Xiong et al. used MD simulations to determine the mechanical properties of monolayer TMDs[31]. They also evaluated the effectiveness of four empirical potential functions, including CVFF1, CVFF2, SW, and REBO potentials, in explaining the mechanical behavior of single layer MoS2 using the benchmark established by DFT calculations[31]. Yang et al. performed both experimental and first principle calculations for understanding the brittle fracture of 2D $MoSe_2$[28]. These computational methods also facilitates the investigation on vertical and lateral heterostructures TMDs[32], [33]. On the other hand, Bertolazzi et al. used an atomic force microscopy (AFM) nano-indentation technique to determine the mechanical properties of exfoliated mono- and bi-layer $MoS_2$[34]. They found that in-plane Young's modulus of monolayer and bi-layer h-$MoS_2$ as 120~240 N/m and 190~330 N/m, respectively[34]. Similar studies on $WS_2$ and $WSe_2$ monolayer and multilayer were also reported[35], [36]. All of these studies were carried out in a controlled environment. On the contrary, the prospect of repeated exposure to extreme temperatures and severe chemical environments during synthesis and practical applications makes monolayer TMDs vulnerable to the evolution and growth of imperfections. Imperfections such as cracks and defects significantly influence the mechanical[37], electrical[38], optical[39], and magnetic[40] properties of 2D TMDs[41]. In case of mechanical properties, it was demonstrated that that multi-atom defects, such as cracks and pores, drastically decreases the fracture strength of TMDs[42]. Moreover, electronic devices made of monolayer TMDs are exposed to various operating conditions. Therefore, it is critical to understand how monolayer TMDs with pre-existed crack respond mechanically in different environments such as under varying temperature. Obtaining material properties experimentally or computationally for various TMDs under different operating conditions is time-consuming and difficult, and in some instances, impossible. In this regard, data-driven a machine learning (ML) based strategy can be very effective.

ML approaches are gaining popularity in the age of big data, especially in materials science[43]. Researchers used ML models in various aspects of material science including material properties prediction[44]–[49], structural topology optimization[50]–[52], design of spinodoid metamaterials[53], multi-scale modeling of porous media[54], design of tessellate composites[55]–[57], architected materials design[58], material properties extraction[59], prediction of hyperelastic or plastic behaviors[60]–[62], process-structure-property linkage[63] and so on. One of the major goals of ML models is to accomplish high-throughput measurement of critical characteristics of materials under different conditions[64]. Previously, ML algorithms such as Support Vector Machine (SVM), and Feed Forward Neural Network (FFNN) have been used to predict mechanical properties of 2D materials [29], [30], [65]–[69]. Wang et al. used SVM for predicting the mechanical properties of $MoSe_2$[29]. Xu et al. investigated four ML models for predicting the mechanical strength of nanocrystalline graphene oxide[67]. Among these four ML models, they found eXtreme Gradient Boosting algorithm is more capable in accurate predictions[67]. Kastuar et al. proposed a ML based toolkit (ElasTool toolkit) for studying the temperature-dependent mechanical properties of 2D $MoS_2$-$WS_2$ heterostructures[68]. The primary constraint of these works is that the proposed ML methods are only applicable to a single 2D material system. Moreover, the effect of defects and cracks are also ignored. Furthermore, these studies are only employed to predict certain mechanical properties such as fracture stress, fracture strain, Young's modulus etc. We need to know beyond these critical mechanical parameters, often times the entire stress strain curve to understand the mechanical response under different strain conditions. Therefore, a generalized ML architecture that is capable of predicting the stress-strain behavior of multiple TMDs with pre-existing cracks is necessary to overcome these limitations.

In this work, we propose a unique Long Short-Term Memory (LSTM) model[70] that predicts the stress-strain response of 2D monolayer TMDs (h-$MoS_2$, h-$MoSe_2$, h-$WS_2$, h-$WSe_2$) with pre-existing cracks under uniaxial tensile tests at various temperatures and chiral orientations. The stress-strain relation of monolayer TMDs is represented by a sequence of consecutive data points, each of which is linked to the preceding stress-strain data points. Because of the temporal connections in the hidden layers of LSTM, this is one of the most suitable ML architectures for predicting stress-strain behavior of TMDs. We also use the FFNN model[71] to predict the critical mechanical properties of the materials. The Feed Forward Neural Network (FFNN) models are widely used in predicting specific material properties due to their lower computational costs and considered as simplest and quintessential ML models[44], [47], [72].

Here, we first obtained the strain-stress relationship for all these TMDs under varied conditions using MD simulations. MD simulations are a well-known computational method for obtaining material characteristics and researchers have previously utilized this technique to generate dataset for ML models. [65], [67], [73] . We utilized the stress-strain data obtained from MD simulations as train data for the LSTM model. Once trained, the LSTM model predicts the full stress-strain response with more than 95% accuracy. We also extracted mechanical properties (fracture stress, fracture strain, and Young's modulus) from the MD simulations, and fit them using a FFNN. The FFNN model is found to predict the mechanical parameters for validation set with 99% accuracy. We compare the predictions of FFNN and LSTM. The results indicate that the LSTM approach has very similar accuracy level as FFNN and is sufficiently capable of predicting the stress-strain behavior of TMD monolayers. Using LSTM model, we can predict the stress-strain profile and other important mechanical properties without conducting time-consuming MD simulations.

## Methodology

**2.1 MD Simulation Method**

In this work, MD simulations are used to conduct uniaxial tensile testing of 2D monolayer TMDs (h-$MoS_2$, h-$MoSe_2$, h-$WS_2$, h-$WSe_2$) in a variety of conditions. The Large-scale Atomic/Molecular Massively Parallel Simulator (LAMMPS)[74], is used to perform all the simulations. Initially, 30 nm × 30 nm nanosheets are created as shown in Figure 1, and cracks with various lengths are introduced in either armchair or zigzag orientations. The cracks are formed by deleting a few rows of atoms, a procedure that has been used in prior studies[75]. The size of the nanosheet was chosen to minimize the effect of sample size while maintaining a reasonable computation time. The fracture lengths are chosen such that the nanosheet width and length are at least ten times larger than half the crack length to eliminate finite dimension effects[75].

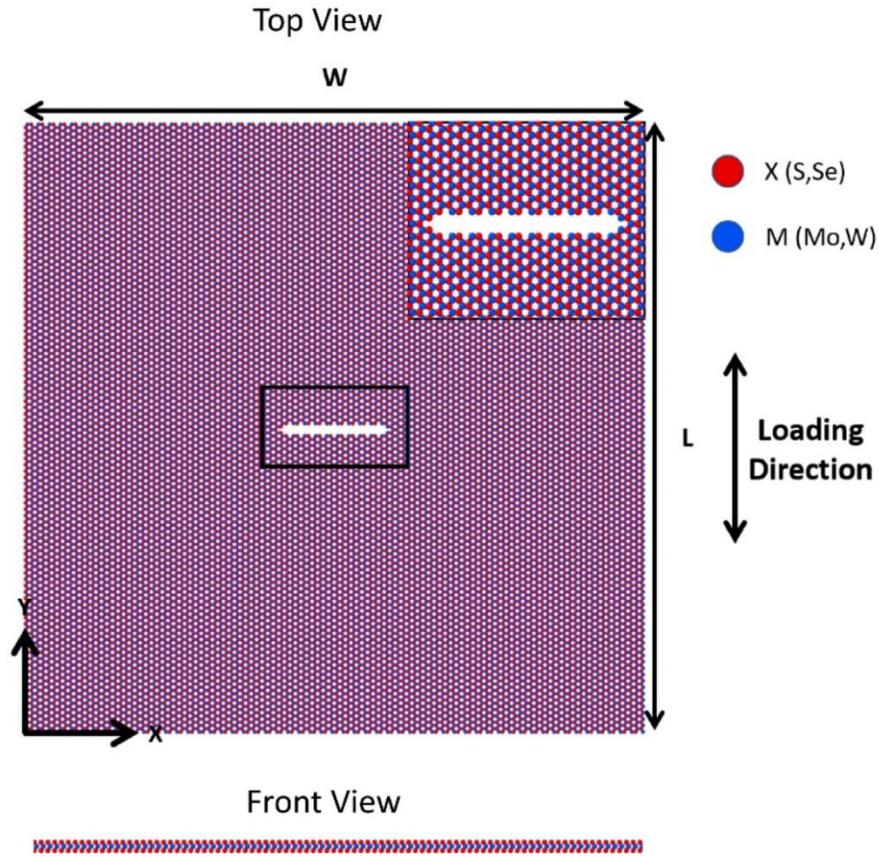

**Figure 1. Initial Structure of TMDs monolayer.** Here, red atoms indicate the Chalcogen, and blue atoms indicate the Metal. L and W denotes the side lengths of the monolayer TMDs. The inset figure showed the crack at the center of the nanosheet, and loading is applied to the perpendicular direction of the crack.

The predictability of MD simulations is dependent on the choice of interatomic potential. The Stillinger–Weber (SW) potential is effective in determining the mechanical characteristics of 2D monolayer TMDs. The SW potential developed by Jiang et al. is employed in this research[76]. This potential has already been used in numerous studies to predict mechanical responses of various layered materials[29], [33]. The atomic interactions in the SW model are expressed as,

$$E = \sum_i \sum_{j>i} \phi_2(r_{ij}) + \sum_i \sum_{j \neq i} \sum_{k>j} \phi_3(r_{ij}, r_{ik}, \theta i_{jk}) \qquad (1)$$

Where $\phi_2$ & $\phi_3$ represent the bond stretch (two-body) and bond bending (three-body) interactions, respectively. To eliminate the effect of free edges, the simulation boundary conditions are specified as periodic in the in-plane (x and y) directions, and the simulation box is assumed to be an exact multiple of the unit cell in these directions. The free boundary condition is applied in the out-of-plane direction to avoid cross-layer interactions.

The system is initially minimized using the Polak-Ribiere conjugate gradient approach[77]. Then the system is equilibrated for $50\ ps$ using the NVE (constant energy, particle) ensemble, followed by an isothermal-isobaric (NPT) ensemble[78] for $100\ ps$ at the operating temperature. Following that, a constant strain rate of $10^9 s^{-1}$ is applied along the in-plane perpendicular direction of the previously created crack using the canonical (NVT) ensemble[78]. We note that the strain rate is significantly higher compared to experiments; however, such a high strain rate has been routinely employed in atomistic scale simulations[75], [79]–[81] to examine material failure events since it enables simulations to be performed with a reasonable computational resource. We computed the atomic stresses using the virial stress formula[82]. A timestep of $1\ fs$ is used for all the simulations. For each case, the average values of the mechanical properties are derived from four separate simulations with various initial conditions. To verify our simulation technique, we compare our results with the previous studies. A comparison of the values is shown in Table S1. The agreement with previously reported data validates the computation approach we used in this study.

**2.2 LSTM Modeling Details**

*2.2.1 LSTM neural network*

The temporal connections of hidden layers between the LSTM units differentiate the LSTM neural network from other neural networks. Typically, a conventional LSTM unit consists of an input gate, an output gate, a forget gate, a unit input, a cell state, and peepholes. These gates act in coordination to direct units to retain or forget particular information derived from temporal input. Figure 2(a) depicts a simplified representation of a typical LSTM unit. The $X_t$ and $y_t$ are used in this research to signify the input and output data, respectively, for the simplified LSTM unit at time $t$. The following calculations are carried out in simplified LSTM units:

$$\begin{aligned}
i^t &= S(\overline{\overline{i^t}}) = S(W_i \times X^t + R_i \times y_L^{t-1} + b_i) \\
o^t &= S(\overline{\overline{o^t}}) = S(W_o \times X^t + R_o \times y_L^{t-1} + b_o) \\
f^t &= S(\overline{\overline{f^t}}) = S(W_f \times X^t + R_f \times y_L^{t-1} + b_f) \\
z^t &= T(\overline{\overline{z^t}}) = S(W_z \times X^t + R_z \times y_L^{t-1} + b_z) \\
c^t &= i^t \odot z^t + c^{t-1} \odot f^t \\
y^t &= T(c^t) \odot o^t
\end{aligned} \qquad (2)$$

Where $W$, $b$, and $R$ denote the weights of input data, biases, and common recurrent weights, respectively. By contrast, the subscripts $z$, $i$, $o$, and $f$ signify the input units, input gate, output gate, and forget gate, respectively. At time $t$, the variables $\overline{\overline{z^t}}, \overline{\overline{i^t}}, \overline{\overline{o^t}}$ and $\overline{\overline{f^t}}$ represent the input data of the input units, input gate, output gate, and forget gate. Similarly, $z^t$, $i^t$, $o^t$ and $f^t$ represent the output data from the corresponding gates at time $t$. The LSTM unit's cell state is denoted by $c^t$. $S$ and $T$ are the sigmoid and hyperbolic tangent activation functions, respectively. The matrix's element-wise product is denoted by $\odot$. The three-dimensional network is displayed in Figure 2(b) and is represented using LSTM units along with the height, width, and time dimensions.

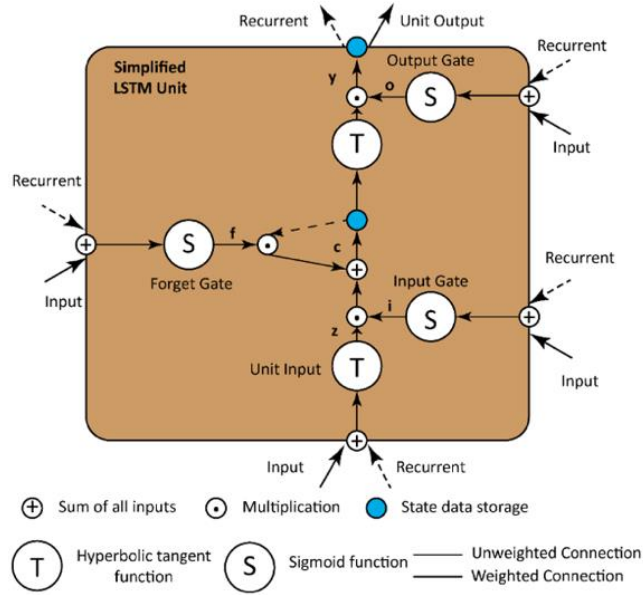

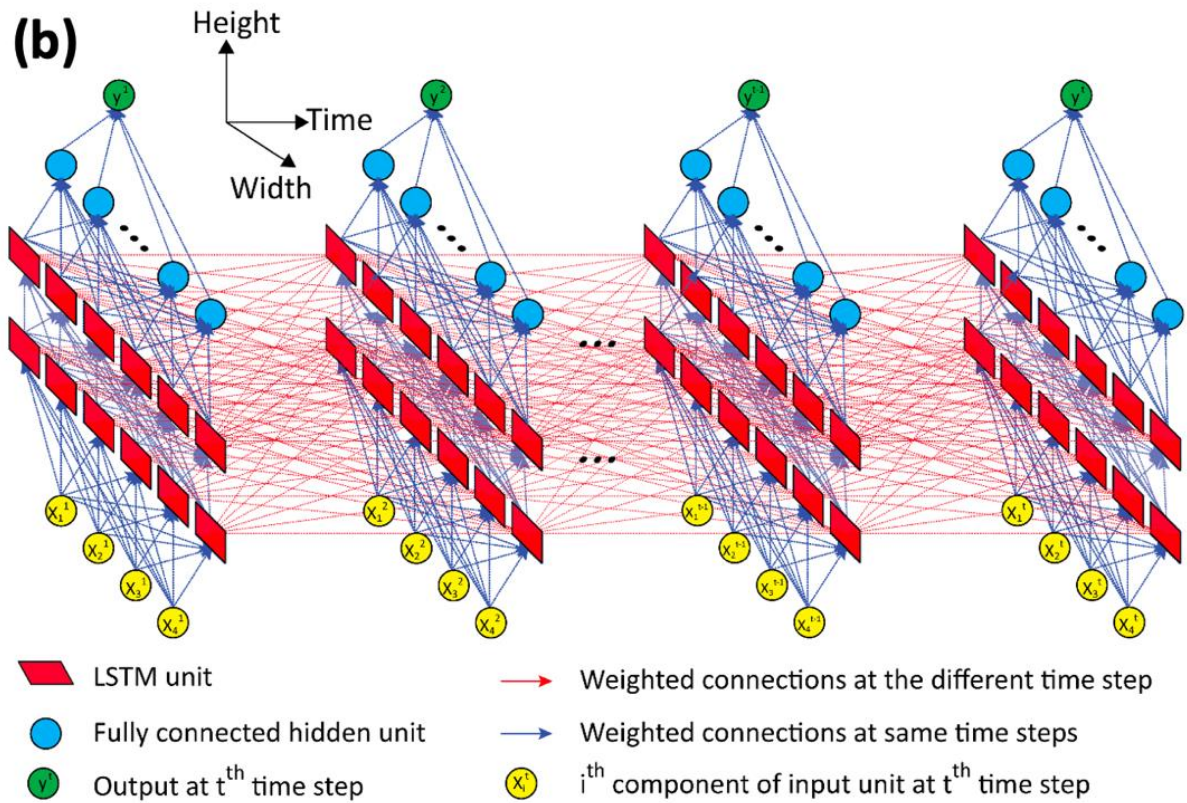

**Figure 2**. (a) Processing of data in a simplified LSTM unit & (b) The LSTM deep learning network's 3D architecture. A simplified LSTM unit is composed of an input gate, forget gate, output gate and unit input. Figure 2(b) denotes the desired LSTM architecture which is comprised of time distributed fully connected hidden layer and the LSTM unit.

*2.2.2 Dataset preparation for LSTM*

Stress-strain data from MD simulations are used as input to train the LSTM model, which then incorporates the TMD's constitutive law into the weight of all connections. For LSTM models, the choice of the appropriate descriptors is essential. The stress-strain relationships of the studied TMDs are examined by varying the chiral orientation, temperature, and pre-crack lengths. In order to describe the TMDs, we employ the metal and chalcogen atomic masses as descriptors. We summarized our input features in Table 1. We studied two chiral directions of 4 materials, and each direction has 5 different crack lengths and 6 different temperatures. Therefore, our dataset contains 300 timesteps in each of 288 data samples.

Table 1. Description of Input features of LSTM model

| Features | Values |
| --- | --- |
| Metal atomic mass | 95.95(Mo),183.84(W) |
| Chalcogen atomic mass | 32.06(S),78.971(Se) |
| Chiral directions | Armchair, Zigzag |
| Temperature | 100K, 200K, 300K, 400K, 500K, 600K |
| Crack length | 1nm, 2nm, 3nm, 4nm, 5nm & Pristine |
| Strain | 0% to 30% |

The values of each variable in this dataset are spanned over a wide range. As a result of their broad ranges, variables with their original values may have greater weight. We scale features of the dataset to address this challenge using the following equation.

$$x' = \frac{x - x_{min}}{x_{max} - x_{min}} \tag{3}$$

Where, $x'$ and $x$ represent the scaled and unscaled values. A system may fail to generate correct results when the input data is not mapped one-to-one, which is a crucial condition for the LSTM technique or any other neural network to perform properly. Our data is organized in a manner that meets the criteria.

*2.2.3 Determination of the LSTM architecture and optimization method*

The LSTM recurrent NN (RNN) was determined empirically in this research. The number of LSTM layers, the number of fully-connected hidden layers, and the number of nodes in each layer describe the LSTM network's architecture. Stress-strain behavior of TMDs can be predicted using time-distributed 5 fully connected hidden layers with varied numbers of nodes in each layer and two LSTM layers for recognizing temporal data. Time-distributed fully connected layers (FCL) are very useful in time-series data. It allows a layer for each input in LSTM/RNN models. Table 2 shows the hyper-parameters of this architecture sequentially.

Table 2. Details of the LSTM architecture.

| Layer name | Number of Nodes | Activation Functions |
|---|---|---|
| Time distributed FCL 1 | 6 | Tanh |
| Time distributed FCL 2 | 30 | Tanh |
| LSTM 1 | 30 | Tanh |
| LSTM 2 | 30 | Tanh |
| Time distributed FCL 3 | 30 | Tanh |
| Time distributed FCL 4 | 6 | Tanh |
| Time distributed FCL 5 (Output) | 1 | Linear |

Gradient descent techniques were utilized to optimize the LSTM network. This optimization technique is especially popular due to its minimal processing cost when dealing with massive datasets. We optimize using stochastic gradients and Adaptive Moment Estimation (Adam). The open-source TensorFlow 2.0 library was used to implement the model.

**2.3 FFNN Modeling Details**

*2.3.1 Artificial Neural Network (ANN)*

Figure 3(a) denotes a simple artificial neuron. From Figure 3(a), the individual inputs $X_1, X_2, X_3, \ldots X_n$ are multiplied by the corresponding weights $W_1, W_2, W_3, \ldots W_n$ and fed up with a summation function which is mainly a dot product of $\boldsymbol{W}.\boldsymbol{X}$ where, $\boldsymbol{X} = [X_1, X_2, X_3, \ldots X_n]^T$ & $\boldsymbol{W} = [W_1, W_2, W_3, \ldots W_n]$. The dot product is summed with a threshold bias $\boldsymbol{b}$ and utilized as an argument in the activation function.

$$Y = \boldsymbol{f}(\boldsymbol{W}.\boldsymbol{X} + \boldsymbol{b}) \tag{4}$$

Where $f$ is the activation function and $Y$ is the output.

A FFNN is a multilayer ANN in which information travels from inputs to outputs without any feedback loops, and neurons in the same layer are not linked to each other but are connected to all neurons in neighboring layers. The architecture of a FFNN is shown in Figure 3(b).

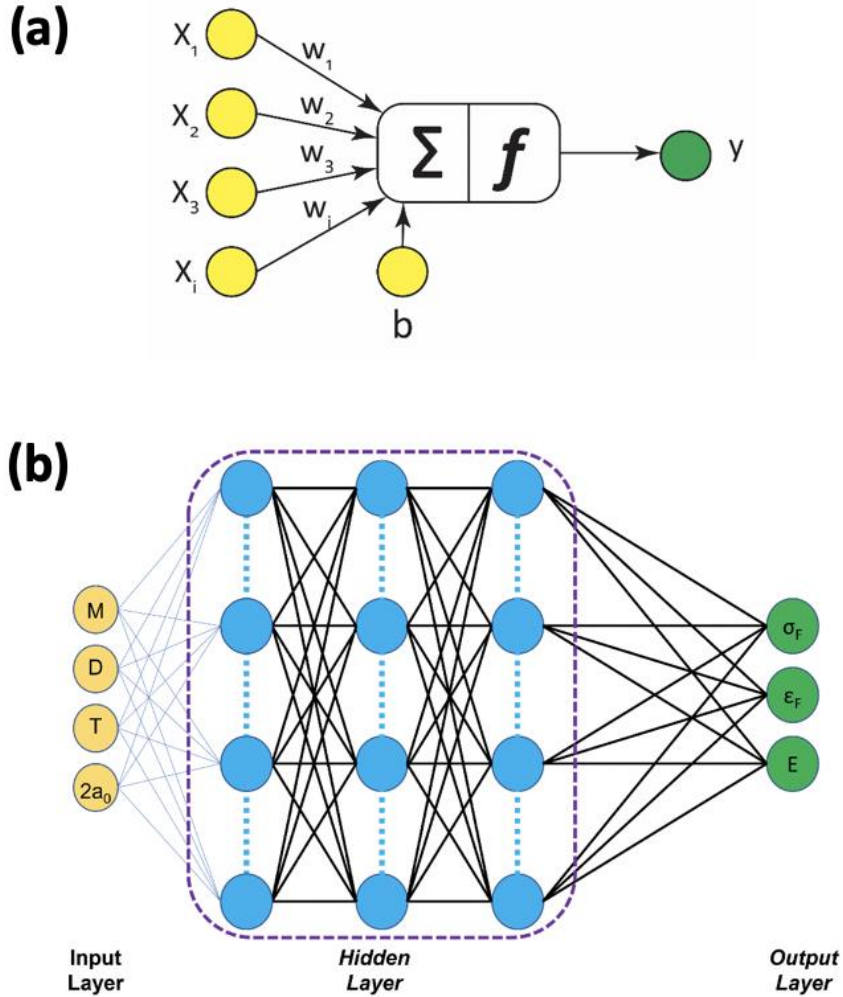

**Figure 3**. (a) Data processing in a simplified FFNN unit and (b) Schematic representation of FFNN. In figure 3(a), X and y indicate the input and output features, respectively. W and b are the weight and bias of a neuron. In figure 3(b), input features are atomic mass ($M$), chiral direction ($D$), system temperature ($T$) and crack length ($2a_0$) and the output features are the fracture stress ($\sigma_F$), fracture strain ($\epsilon_F$) and Young's modulus (E).

**2.3.2 Dataset preparation for FFNN**

The first step in developing any ANN model is to create a dataset. We used data from the tensile test simulations of monolayer TMDs carried out on four distinct TMDs in varying chiral directions, six operating temperatures, and six different crack lengths, as well as on their pristine structures. We conducted 5 separate simulations for each case with different random initial velocity distribution that resulting in a total of 1440 simulations. We used the median values for each set of output data. We have five input and three output parameters in our dataset. The input parameters include atomic masses of metals and chalcogens, chiral direction, temperature, and the initial crack length. Table 3 shows the input parameters and their values. The output data includes the fracture stress, fracture strain, and Young's modulus. Physical background and previous experience are used to identify and decide both property-related outputs and input attributes. The fracture stress and strain are determined by taking the stress-strain value from the stress-strain curve at fracture point. Young's modulus is calculated by fitting stress-strain data up to fracture with two degrees polynomial and using the co-efficient of the first-degree term[83]. We use z-score normalization to scale the features of the dataset in order to minimize the influence of different ranges, which may have varying weights.

Table 3. Description of input features of FFNN model

| Features | Values |
|---|---|
| Metal Atomic Mass | 95.95(Mo),183.84(W) |
| Chalcogen Atomic Mass | 32.06(S),78.971(Se) |
| Chiral Directions | Armchair, Zigzag |
| Temperature | 100K, 200K, 300K, 400K, 500K, 600K |
| Crack Length | 1nm, 2nm, 3nm, 4nm, 5nm & Pristine |

*2.3.4 Architecture Determination and Optimization method*

To develop the NN model, we need to select hyper-parameters, the number of hidden layers, the number of neurons in each layer, and the initial values of connection weights and activation functions. To determine the optimal structure of the hidden layer, a numerical optimization technique based on a global search across the whole accessible space is applied. The accessible space is defined in this research as any NNs with 3 to 5 hidden layers each comprising of 50 to 65 neurons. This search space was found to be sufficiently big to discover the optimum NN for this application. The hyper-parameters of this design are listed consecutively in Table 4.

Table 4. Details of FFNN architecture.

| Layer name | Number of Nodes | Activation Functions |
|---|---|---|
| Input Layer | 5 | - |
| Fully connected layer (FCL) 1 | 50 | Relu |
| FCL 2 | 50 | Relu |
| FCL 3 | 50 | Relu |
| (Output Layer) FCL 4 | 3 | Linear |

In this FFNN model, the backpropagation model is used to calculate the derivatives of the error function for the hidden layer weights and biases. To decrease the error functions, batch gradient descent methods were used. The Batch gradient method determines the gradient of the entire dataset for each iteration. Stochastic gradient descent (SGD) is used as an optimizer in this study.

**2.4 Error Analysis and Evaluation Indicators**

We utilized a typical loss function for the LSTM and FFNN deep learning methods. The loss function was made up of the MSE function and the $L_2$ regularisation term, which is specified as follows:

$$Loss = \frac{1}{N} \sum_{i=1}^{N} \frac{(y_{pi} - y_i)^2}{2} + \frac{\lambda}{2N} \times \left( \sum W^2 \right) \tag{5}$$

To assess the performance of different models, the coefficient of determination ($R^2$) was utilized. A larger coefficient of determination signifies a more accurate prediction. The following definition applies to the variable $R^2$:

$$R^2 = \frac{\sum_{i=1}^{N}(y_i - y_{pi})^2}{(y_i - \bar{y})^2} \tag{6}$$

Here, the targeted data for each set of samples is represented by $y_i$, the output by $y_{pi}$, and the average by $\bar{y}$. N is the number of total sample points. The $L_2$ regularization parameter $\lambda$ prevents overfitting by penalising the weights ($W$).

## 3. Results and Discussion

### 3.1 LSTM Model

The adjusting of hyper-parameters affects the accuracy of any deep learning model. [84] There are several hyperparameters to consider while developing a deep learning model, including activation functions, batch size, and learning rate, all of which need to be optimized to fine-tune each layer. These parameters influence other model parameters such as weights and biases and, as a result, their performance is influenced by the values of hyperparameters. Finer tweaking of these hyperparameters almost always leads to increased model accuracy. We have investigated an optimal technique based on minimizing validation loss in this work. Figure 4 depicts the losses in the training process for training and validation samples. The training and validation losses are found to decrease exponentially with increasing number of iterations and eventually become steady. After approximately 400 iterations, the validation loss approaches the training loss and becomes persistent with an MSE value of ca. 0.002. As seen in Figure 4, our LSTM model converged optimally, with neither over- nor under-fitting. This indicates that the LSTM model is a acceptable model for predicting the mechanical behavior of monolayer TMDs.

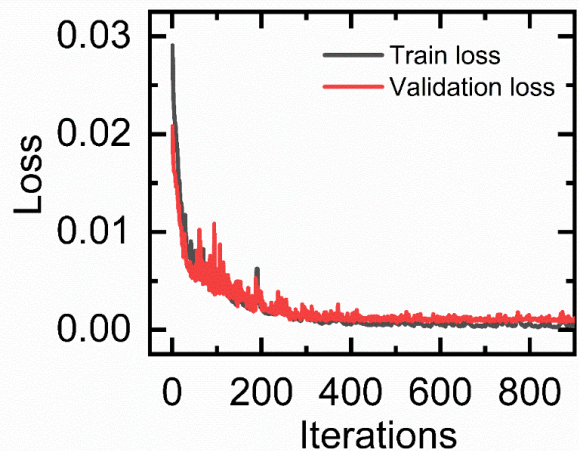

**Figure 4. Loss in the training process in the LSTM model.** The training and validation loss decrease exponentially with the increasing number of iterations.

Figure 5 presents the comparison of the $R^2$ value among the train, validation, and test samples. The $R^2$ value is calculated from the actual and predicted stress-strain diagram by using Equation. 6. Here, the train samples are used to fit the model and the validation samples are used to evaluate the model. Test samples are unknown to our model. From Figure 5, the average $R^2$s in this LSTM model is close to 1 for train and validation samples and more than 95 percent for unknown test samples. Clearly, the result indicates an excellent fit between the ML predicted stress-strain curves and those produced from MD results, for all samples. This indicates that the LSTM-based approach employed in this study is highly accurate in predicting the stress-strain behaviour of 2D TMDs.

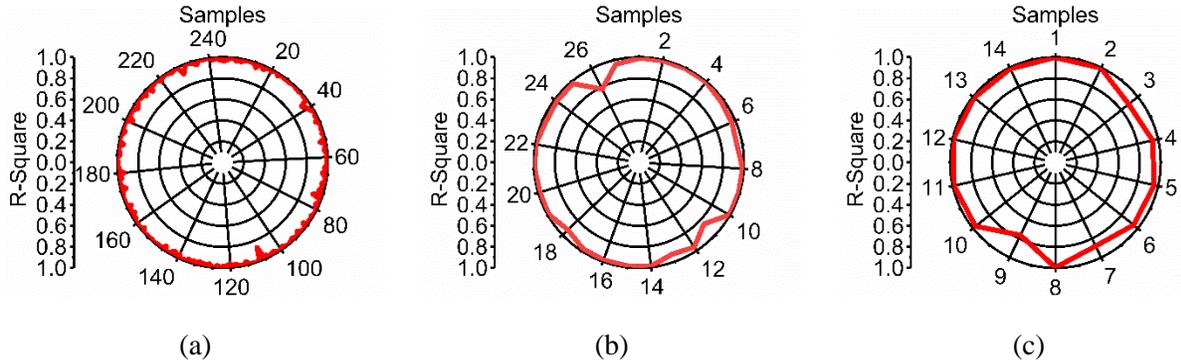

(a) (b) (c)

**Figure 5.** Coefficients of determination of different samples for the LSTM model: (a) training samples, (b) validation samples (c) testing samples.

Next, we aim to understand how the LSTM model performs in predicting the entire stress-strain curve of the TMDs considered in this study. Figure 6 compares the stress-strain profile predicted by the LSTM model to the curve obtained by MD simulations for 8 representative testing samples. In each of the subfigures, the red curves are the predictions from the LSTM model whereas the black curves are obtained from MD simulations. These 8 samples are chosen from the 14 test samples to represent all four materials and two chiral orientations. However, the crack length and temperature conditions were chosen randomly. It is evident from Figure 6 that stress-strain curves predicted by the LSTM model nicely coincide with the curves obtained from MD simulations. This indicates the high degree of reliability of the LSTM model employed in this study in predicting the mechanical response of the TMDs. We also provide a similar comparison for training and validation samples (See Figure S1 and Figure S2). It is clear from Figures S1 and S2 that the model has even higher accuracy for training and validation samples. Since it is capable of predicting the full stress-strain response, it is possible to extract the important fracture parameters (fracture stress and fracture strain) from this model and compare them with the obtained values from MD simulations.

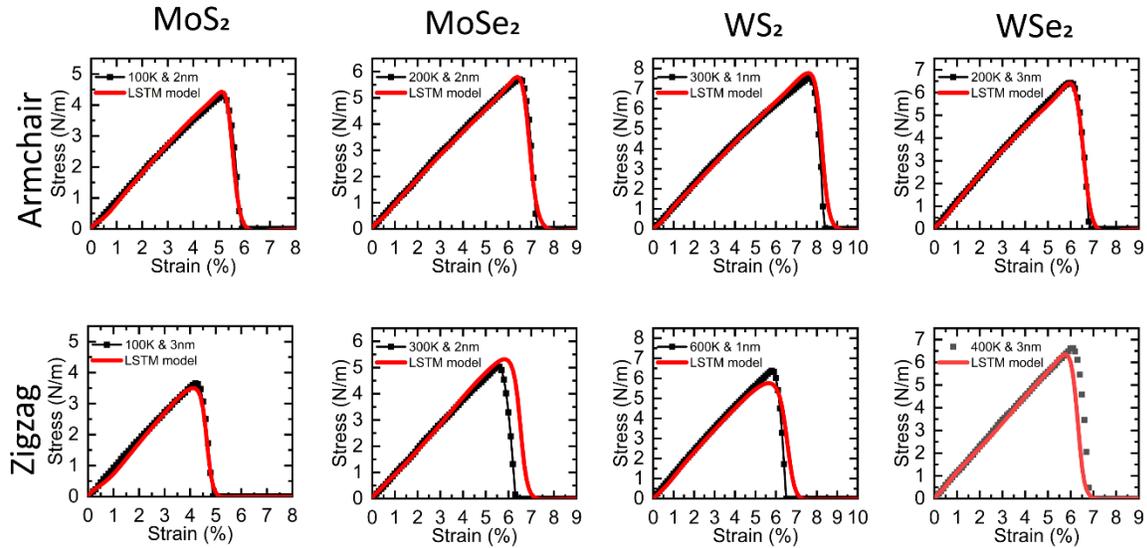

**Figure 6.** Predictions of LSTM models for testing data samples.

Next, we extracted the fracture stress and strain data using the LSTM model's stress-strain prediction and compared them to the values obtained from MD simulations. Figure 7(a) shows the fracture stress of TMDs obtained from the LSTM predicted stress-strain curve along with the data obtained from MD simulations for all the training, validation, and testing samples. The curve shows a remarkable accuracy with the coefficient of determination as accurate as 0.995. We also plot a similar curve for fracture strain in Figure 7(b). The fracture strain predictions using LSTM are also highly accurate, with a coefficient of determination of 0.999.

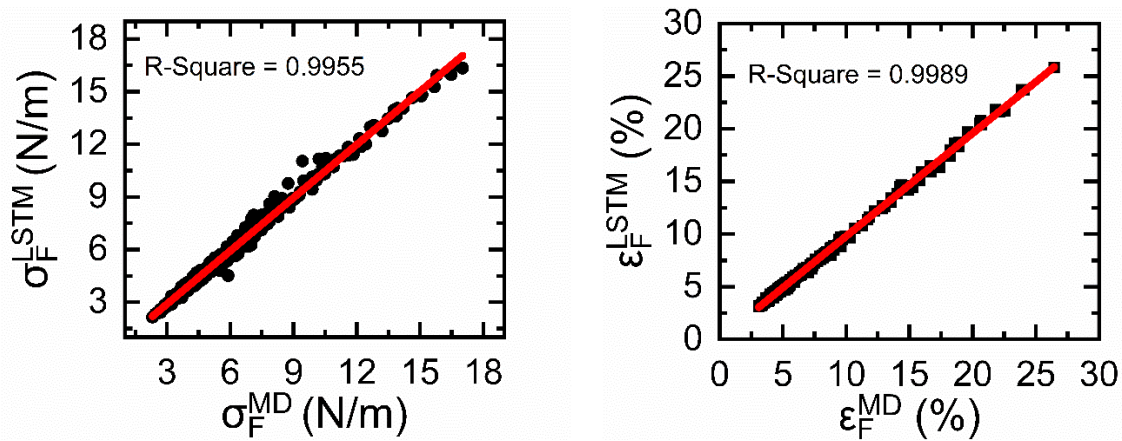

(a) (b)

**Figure 7.** Comparison of predicted results from the LSTM model and results obtained from MD simulations (a) Fracture stress, (b) Fracture strain.

The high accuracy of the LSTM model, as depicted in this study, implies that we can predict the mechanical behaviour of TMDs with a high degree of reliability without performing computationally intensive simulations. Here, we used MD simulation data as training samples. We expect that our model will be equally accurate for experimental training data as well. Here we only presented the accuracy of LSTM for four TMDs; however, we anticipate that the model should be applicable to various other TMDs as well as other 2D materials if the appropriate descriptors are chosen.

**3.2 FFNN Model**

Next, we use the FFNN model to predict the fracture properties of the four TMDs. Figure 8 depicts the loss of the train set and validation set throughout the FFNN training process. Both the training and validation losses drop exponentially as the number of iterations increases and converge around the MSE value of 0.1. Figure 8 also illustrates that the FFNN model is well-fitting, with the train and validation loss curves overlapping each other.

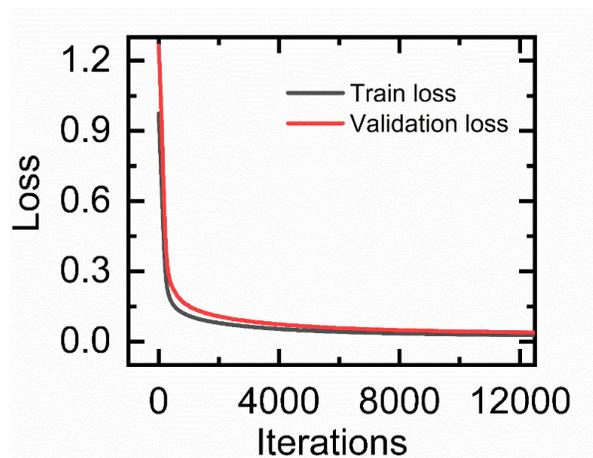

**Figure 8. Loss in the training process in the FFNN model.** The training and validation loss decrease exponentially with the increasing number of iterations.

We assess the accuracy of the FFNN model by comparing the results obtained in MD simulations to the FFNN predictions for the training, validation, and testing samples, as shown in Figure 9. We investigated the predictions of three output parameters such as fracture stress, fracture strain, and Young's modulus. A comparison of fracture stress is depicted in Figure 9(a). We calculate the coefficient of determination to be 0.998, which indicates a remarkable accuracy. We also plot the FFNN predicted vs. obtained values of fracture strain and Young's modulus in Figure 9(b) and 9(c), respectively. The accuracy of the prediction of fracture strain and Young's modulus is also very high with a coefficient of determination of 0.994 and 0.997, respectively.

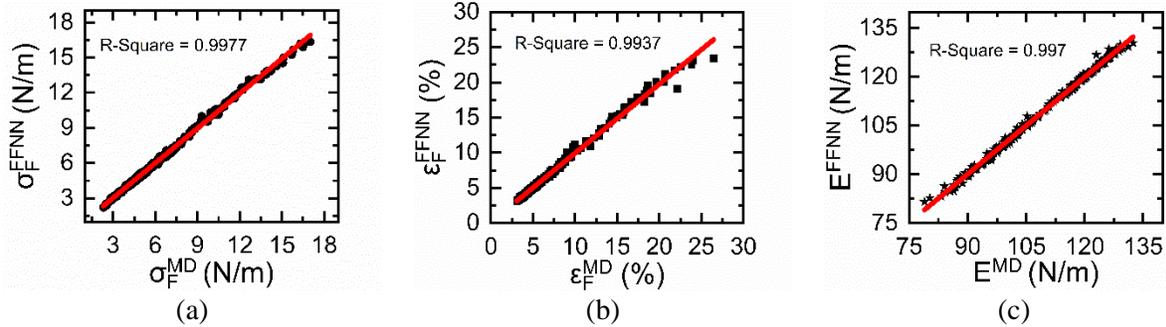

(a)            (b)            (c)

**Figure 9.** Comparison of predicted results from FFNN model and results obtained from MD simulations (a) Fracture stress, (b) Fracture strain, & (c) Young's Modulus.

The FFNN model is highly accurate in predicting the critical mechanical parameters. However, this model is unable to predict the entire stress-strain behaviour accurately as there is no temporal linkage of hidden layers. For this reason, the FFNN model ignores the relationship between each data points of stress-strain curve and instead treats them as separate data points. In many cases, the entire stress-strain response of a material is necessary to assess the materials performance. Therefore, it is to be noted that the FFNN model is only advantageous if one is to determine the mechanical properties of a material.

**3.4 Comparison between LSTM and FFNN**

We further compared the prediction of LSTM and FFNN models to the MD simulations data as presented in Figures 7 and 9. The accuracy of both the models is close to one. However, we can predict the stress-strain behaviour of pre-cracked TMDs using the LSTM model across a wide range of operating temperatures and loading orientations without compromising the accuracy of the predictions. However, it should also be noted that the LSTM model is considerably more complex than the FFNN model. One of the ways to determine the complexity of a neural network is to calculate the number of trainable weights. As the number of hidden layers and nodes in the models increases, the number of trainable weights in NNs increases as well. The following formulae are used to determine the number of trainable weights in the LSTM and FFNN models used in the study:

$$Num_{FF} = \sum_{i}^{L-1} N_i * N_{i+1} + \sum_{i}^{L-1} N_{i+1}$$

$$Num_{LSTM} = \sum_{l}^{Others} N_i * N_{i+1} + \sum_{i}^{Others} N_{i+1} + 4 * (N_{aheadLSTM} * N_{LSTM} + N_{LSTM}^2 + N_{LSTM})$$

(7)

Where $Num_{FF}$ & $Num_{LSTM}$ denote the total number of trainable weights in the FFNN and LSTM neural networks, respectively. $N_i$ denotes the number of nodes in the $i$th layer, $N_{aheadLSTM}$ is the number of nodes in the layer before the LSTM layer, $Num_{LSTM}$ denotes the number of nodes in the LSTM layer, and Others denote the number of nodes in all other levels in the LSTM model except the LSTM layer.

In our FFNN and LSTM NNs, there were a total of 5553 and 16015 trainable weights, respectively. Due to the presence of four recurrent gate units in the LSTM units, the trainable weights in the LSTM model are much larger than those in conventional NNs. Each iteration of the LSTM model incurs a higher compute cost due to the higher number of trainable weights and error backpropagation along the expanded time dimension. Because of a large number of trainable weights and its complicated architecture compared to the FFNN model, the LSTM model needs a larger amount of computational resource. However, by leveraging a graphics processing unit (GPU), the computational complexity of the LSTM model can be further lowered, allowing the LSTM to be used for big-data analysis. When compared to traditional NNs, the enhanced accuracy and convergence rate of the LSTM model justifies the higher processing costs.

## 4. Conclusion

In this work, we developed two ML models: LSTM and FFNN to predict the mechanical behaviour of 2D TMDs. LSTM predicts the entire stress-strain response with an accuracy close to 1 for training and validation samples and 0.95 for unknown test samples. FFNN predicts the fracture stress, fracture strain, and Young's modulus of the materials with a similar level of accuracy. We also found the LSTM approach to be slightly more computationally costly than the FFNN due to its added complexity arising from a large number of trainable weights. However, the ability of LSTM to predict the entire stress-strain response would make it advantageous over FFNN. We used these ML models to determine the mechanical properties of 2D TMDs over a large design space which is impractical to investigate through simulations or experiments. We expect that this ML based approach can also be generalized to predict the other important optical, electrical, thermal, and magnetic properties of 2D materials in general by incorporating the appropriate descriptors.

## Acknowledgment

The authors would like to acknowledge Multiscale Mechanical Modeling and Research Networks (MMMRN) for their technical assistance to conduct the research. We acknowledge the contribution of Arun HPCC (Rajshahi University) for providing computational supports. MMI acknowledges start-up funds from Wayne State University.

## CRediT authorship contribution statement

**Prottay Malakar:** Conceptualization, Data curation, Formal analysis, Investigation, Methodology, Software, Visualization, Writing – original draft. **Md Shajedul Hoque Thakur:** Investigation, Software, Writing – review & editing. **Shahriar Muhammad Nahid:** Conceptualization, Investigation, Writing – review & editing. **Md Mahbubul Islam:** Funding acquisition, Supervision, Writing – review & editing.

## Conflict of Interest

There are no conflicts to declare.

# Supplementary Information of

# Data-Driven Machine Learning to Predict Mechanical Properties of Monolayer TMDs

Prottay Malakar[1], Md Shajedul Hoque Thakur[2], Shahriar Muhammad Nahid[3] and Md Mahbubul Islam[4*]

[1]*Department of Mechanical Engineering, Bangladesh University of Engineering & Technology (BUET), Dhaka, 1000, Bangladesh*

[2]*Department of Materials Science and NanoEngineering, Rice University, Houston, TX 77005, United States*

[3]*Department of Mechanical Science and Engineering, University of Illinois Urbana Champaign, Urbana, IL 61801, United States*

[4]*Department of Mechanical Engineering, Wayne State University, 5050 Anthony Wayne Drive, Detroit, MI 48202, United States*

[*]Corresponding author, *Email address:* gy5553@wayne.edu

| Table S1. Comparison of Young's modulus obtained by our MD simulations and the value obtained in the literature | | | | |
|---|---|---|---|---|
| Material | Armchair Direction | | Zigzag Direction | |
|  | Calculated (N/m) | Literature (N/m) | Calculated (N/m) | Literature (N/m) |
| $MoSe_2$ | 107.40 | 95-110[1] | 105.61 | 95-110[1] |
| $WS_2$ | 126.60 | 115-125[2] | 123.83 | 115-125[2] |
| $WSe_2$ | 129.98 | 119.73[3] | 126.10 | 118.29[3] |

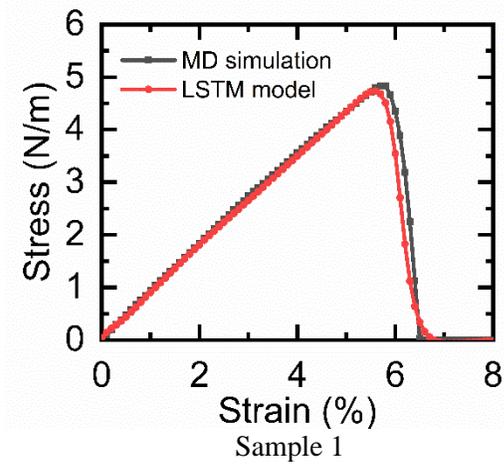
Sample 1

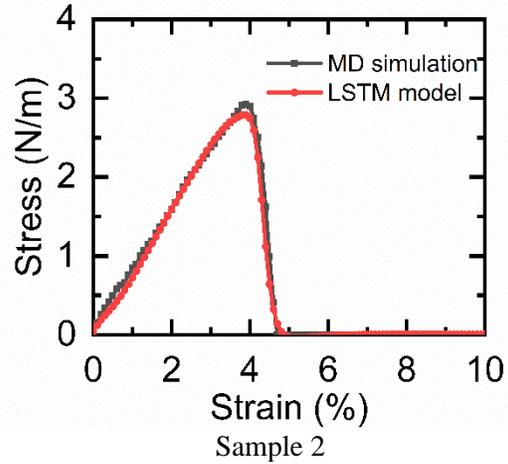
Sample 2

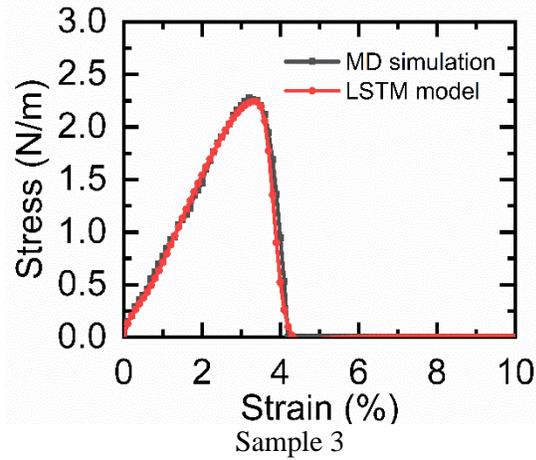
Sample 3

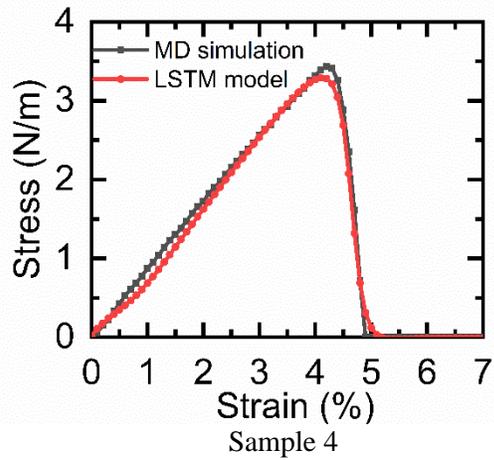
Sample 4

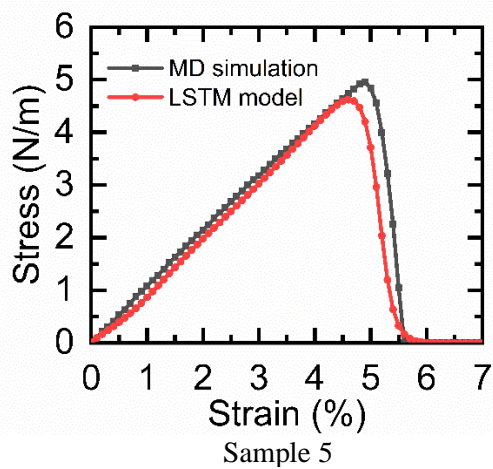
Sample 5

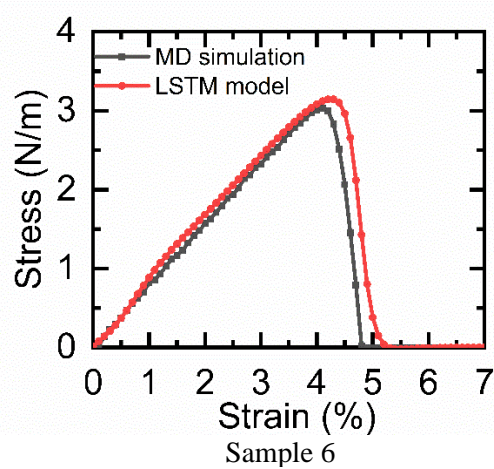
Sample 6

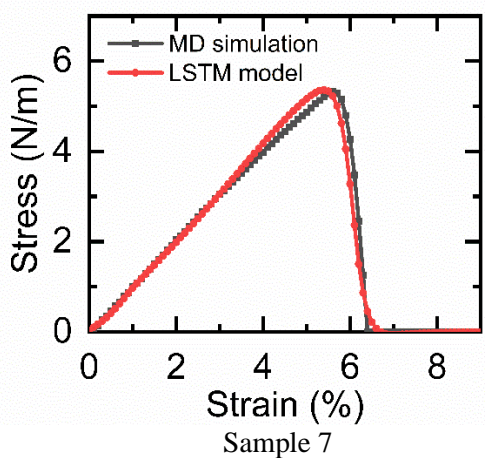
Sample 7

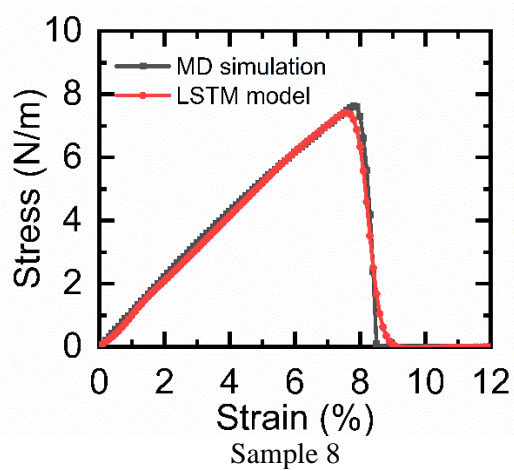
Sample 8

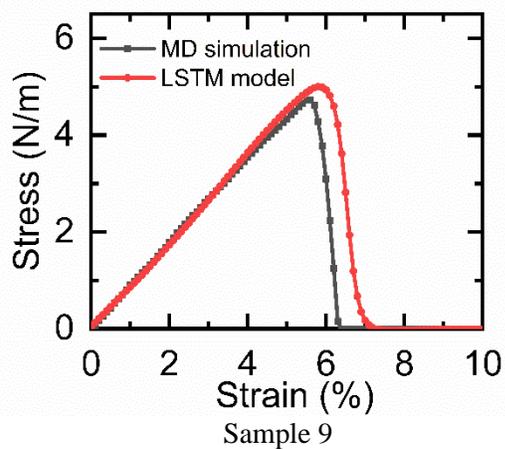
Sample 9

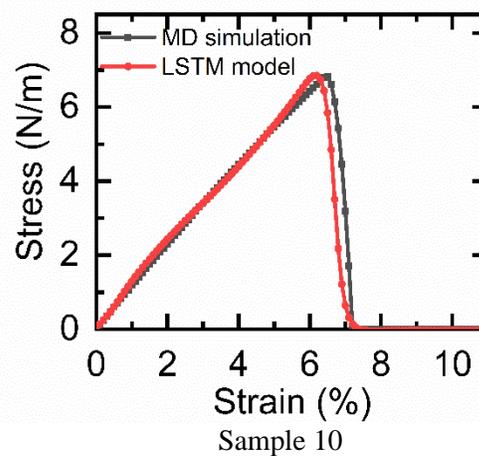
Sample 10

Figure S1. Predictions of LSTM models for training data samples. A few representative cases are shown here.

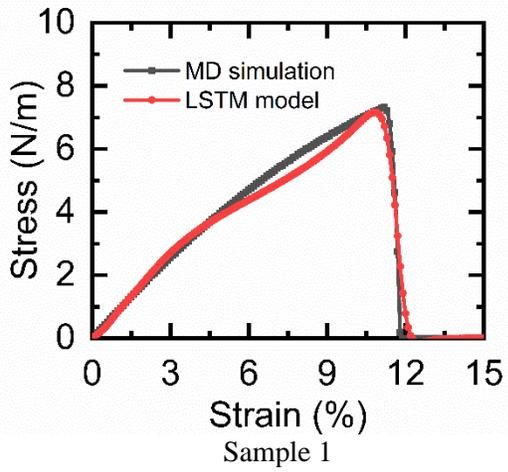

Sample 1

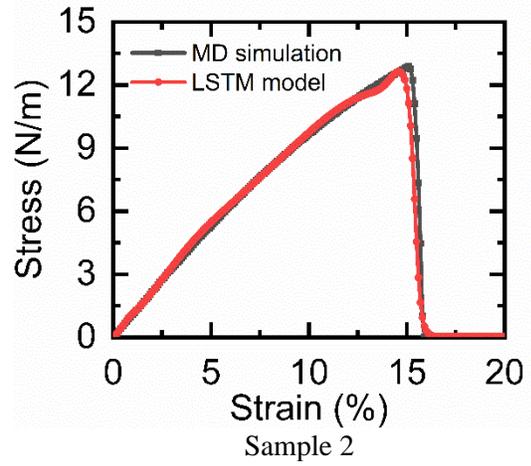

Sample 2

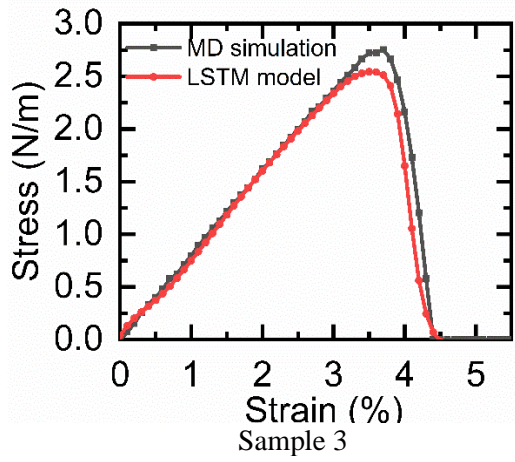

Sample 3

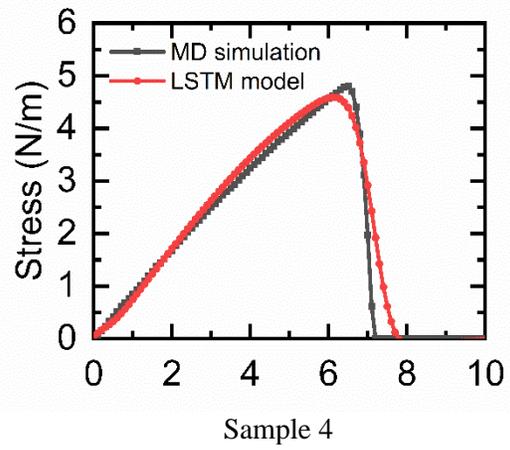

Sample 4

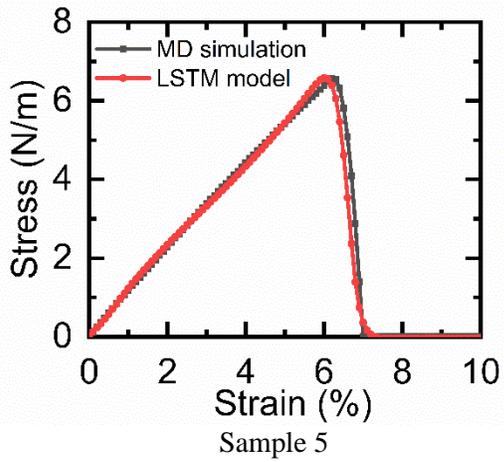

Sample 5

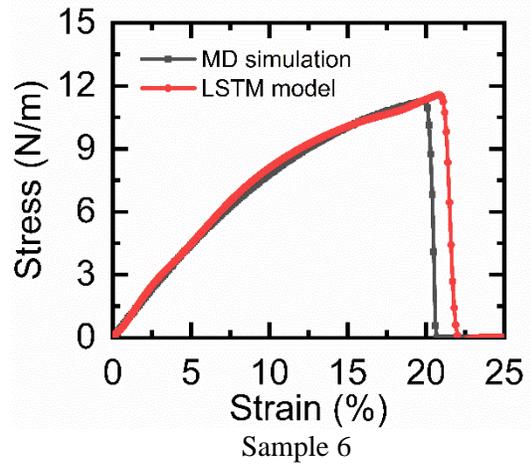

Sample 6

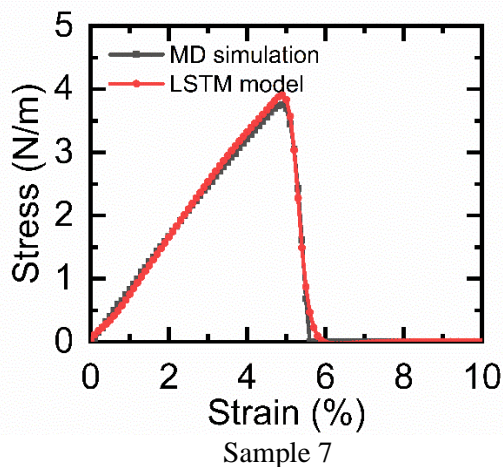
Sample 7

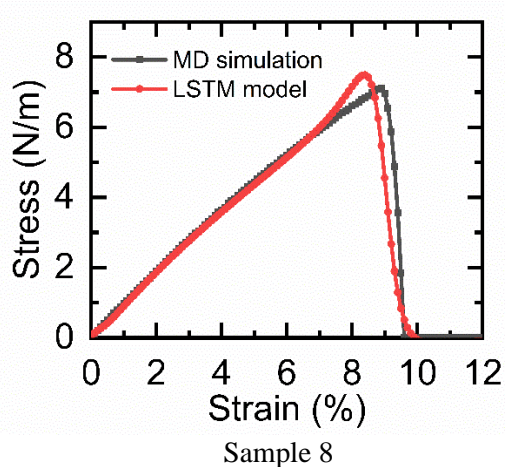
Sample 8

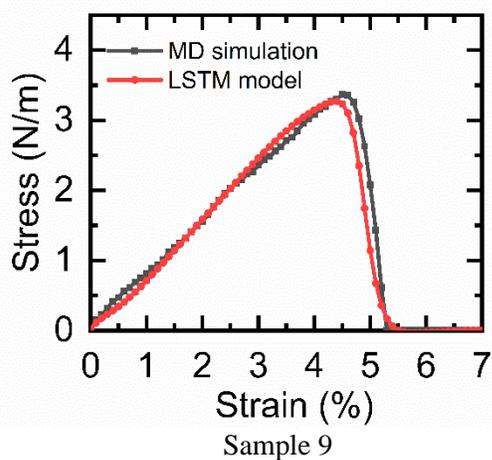
Sample 9

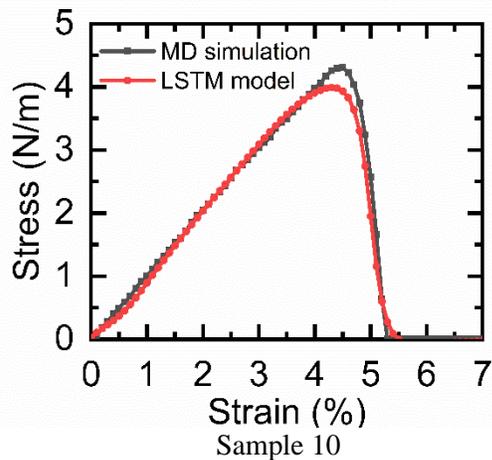
Sample 10

Figure S2. Predictions of LSTM models for Validation data samples. A few representative cases are shown here.